# Low-Excitation Vertical Ion Shuttling in Scalable Multi-Rail Ion Trap Architectures

Qirat Iqbal
*Institute of Physics, University of Sindh*
Jamshoro, Pakistan
qirat.iqbal@scholars.usindh.edu.pk

Altaf H. Nizamani[1]
*Institute of Physics, University of Sindh*
Jamshoro, Pakistan
altaf.nizamani@usindh.edu.pk

***Abstract*: We investigate optimized vertical ion-shuttling protocols for trapped-ion applications across a range of ion-trap experiments, including three-dimensional gradient-measurement sensors, on-chip ion fluorescence collection and imaging, improved laser accessibility, and quantum information processing. In this work, we focus on minimizing motional energy gain during ion transport. Our findings indicate that anomalous heating becomes the dominant limiting factor only for shuttling durations exceeding 500µs, whereas the final motional excitation is strongly dependent on the selected shuttling protocol. Using a recently measured heating rate of (3.1 ± 0.35) quanta ms$^{-1}$ at an ion–surface separation of 134 ± 1.5 µm, we demonstrate that the motional excitation can be restricted to fewer than eight quanta when the ion is vertically displaced to 86µm from its initial position at 134µm within 500µs. These results establish the feasibility of near-adiabatic vertical ion shuttling compatible with the operational requirements of high-fidelity quantum sensing and scalable quantum information processing applications.**

## I. INTRODUCTION

Over the past decade, significant efforts have been devoted to the development of scalable architectures for quantum information processing based on trapped ions. Microfabricated surface-electrode ion traps, in particular, have emerged as a promising platform that enables ions to be confined above a planar electrode surface and transported across complex trapping arrays [1–5]. A central feature of these architectures is the partitioning of ions into distinct trapping zones, where multi-qubit quantum operations are implemented by dynamically merging selected ions within designated processing, computational and read-out regions. Within this framework, ion shuttling plays a critical role in enabling controlled interaction between spatially separated qubits. Reliable ion shuttling has been experimentally demonstrated in linear trap arrays as well as through junction structures, including T-, X- and Y-junction geometries [6–9]. More recently, the successful transport of ions between separate trap modules has been achieved [10], representing an important milestone toward modular and large-scale trapped-ion quantum computers. Parallel efforts have focused on optimizing electrode geometries to minimize motional excitation during transport [11–13].

While axial (linear) shuttle has been extensively explored for ion transport between memory, interaction, and detection zones, vertical ion shuttling or transport perpendicular to the trap surface, has recently attracted increasing attention due to its unique capabilities in surface-electrode trap architectures. More recently, vertical shuttling has also been utilized to investigate anomalous heating in surface ion traps by enabling systematic measurements of motional heating rates as a function of ion–surface distance [14].Vertical shuttling, along with linear shuttling, provides an additional degree of freedom for tailoring the local trapping environment and optimizing ion–field interactions.

Such controlled out-of-plane motion offers several potential applications. By varying the ion–surface distance, vertically shuttled ions can probe spatial variations in electric and magnetic fields with high sensitivity, enabling gradient field measurement experiments and facilitating three-dimensional field mapping for quantum sensing applications. In addition, vertical positioning of ions relative to integrated on-chip optical elements, such as microfabricated mirrors, lenses, or waveguides, can enhance fluorescence collection efficiency, thereby improving state detection fidelity in scalable trapped-ion systems [15–16].

Furthermore, the motional energy gain incurred during ion transport remains a critical performance metric for practical implementations of vertical shuttling. Previous studies by Hucul *et al.* [17] and Nizamani *et al.* [13] have quantitatively examined the excitation induced during transportation and ion separation operations. Building upon these foundational efforts, the present work focuses on the analysis and optimization of vertical ion shuttling protocols in surface-electrode trap arrays. By synthesizing insights from prior experimental and theoretical investigations, we identify key strategies for achieving efficient and adiabatic vertical transport while minimizing motional excitation, thereby supporting high-fidelity quantum sensing and scalable quantum information processing applications.

[1] Corresponding author: Prof Dr. Altaf H Nizamani altaf.nizamani@usindh.edu.pk

In this work, we present a quantitatively optimized vertical ion-shuttling protocol in a surface-electrode trap, based on coordinated RF and DC control of the trapping potential. A key contribution is the systematic analysis of motional excitation as a function of transport parameters, particularly the hyperbolic tangent trajectory, enabling identification of optimal operating regimes that minimize total motional quanta. Furthermore, by explicitly accounting for the interplay between shuttling-induced excitation and anomalous heating, we establish practical conditions for fast, low-excitation transport compatible with scalable trapped-ion architectures.

## II. Trap Modelling

The electrostatic potential of the surface-electrode ion trap was simulated using the analytical formalism developed by House [12]. The calculations were performed in Mathematica within the framework of the gapless-plane approximation. In this approach, individual electrode segments were modeled independently, with each electrode defined by its corner coordinates in the two-dimensional (x, z) plane. The resulting potential contribution of each electrode was then computed using the analytical expressions derived in [12]. The governing equation employed for this modeling is given below.

$$\begin{aligned}\phi\,(x,y,z) = \frac{V_{rf}}{2\pi}\Bigg\{ & \mathrm{Arctan}\left[\frac{(x_2-x)(z_2-z)}{y\sqrt{y^2+(x_2-x)^2+(z_2-z)^2}}\right] \\ & - \mathrm{Arctan}\left[\frac{(x_1-x)(z_2-z)}{y\sqrt{y^2+(x_1-x)^2+(z_2-z)^2}}\right] \\ & - \mathrm{Arctan}\left[\frac{(x_2-x)(z_1-Z)}{y\sqrt{y^2+(x_2-x)^2+(z_1-z)^2}}\right] \\ & + \mathrm{Arctan}\left[\frac{(x_1-x)(z_1-z)}{y\sqrt{y^2+(x_1-x)^2+(z_1-z)^2}}\right]\Bigg\} \qquad (1)\end{aligned}$$

The electric potential $\phi$ generated above the electrode surface by an applied radio-frequency (RF) voltage $V_{RF}$ is described by the analytical expression given in Eq.1. In this formulation, $x_1$ and $x_2$ denote the corner positions of an individual electrode along the x-direction and therefore define its lateral width. Likewise, $z_1$ and $z_2$ specify the electrode extent along the z-direction, corresponding to its longitudinal length. This expression provides a quantitative description of the electric field distribution in the vicinity of the electrode surface when subjected to the applied RF voltage $V_{RF}$.

Using this formalism, we developed a scalable multi-rail surface-electrode trap architecture comprising four symmetrically arranged RF electrodes separated by grounded electrodes to achieve precise radial field shaping, while segmented DC control electrodes distributed around the RF rails generate independently addressable multi-zone trapping potentials along the axial direction, thereby enabling controlled confinement and manipulation of ions within complex trapping arrays in two dimensions.

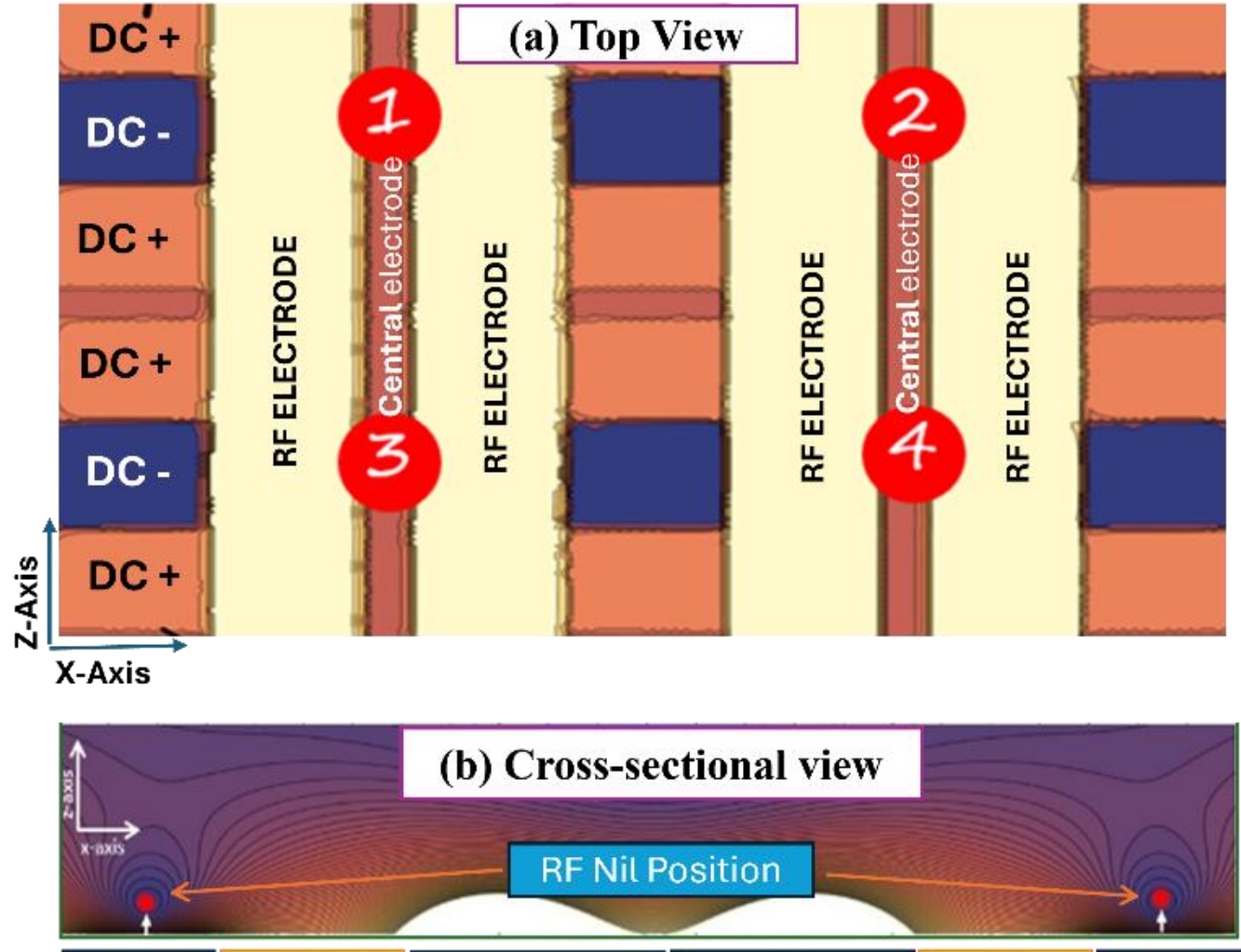

Figure 1. (a) Top view of the optimized multi-rail, multi-region surface-electrode ion trap design. (b) Pseudopotential distribution for two trapping regions, illustrating the formation of local minima at the RF nil position arising from the applied DC and $V_{RF}$ voltages.

The RF electrodes are designed to be comparatively wide along the x-axis and significantly elongated along the z-axis, a geometry that supports extended trapping regions. Fig.1 illustrates the multi-rail surface-electrode on trap comprising four distinct trapping regions. Further details of trap design can be found in [18]. Each trapping zone is located above the trap surface (along the y-direction), as shown in Fig.1(a), and can confine one or more ions simultaneously. It is important to note that a trap composed of RF electrode rails provide confinement in the radial directions (x and y), while axial confinement is achieved through the application of appropriate DC potentials through segmented DC-electrodes. After obtaining the electrostatic field distribution above the electrode surface, the RF trapping potential was calculated using the pseudopotential approximation. This was carried out analytically within the gapless-plane approach. The resulting potential minimum is located above the grounded electrodes, as shown Fig.1(b) from the cross-sectional view of trapping zones 1 and 2, similar can be shown at zone 3 and 4, indicating uniform confinement parameters across the multi-rail surface structure. Further to this,the trap geometry is optimized based on the design principles reported in [17]. As an initial step in the optimization procedure, key geometric parameters were determined, most notably the distance between the trapped ion and the electrode surface, commonly referred to as the ion height '*h*'. In practical surface-electrode ion trap implementations, the ion height is typically chosen to exceed 120μm to ensure sufficient optical access for laser beams within the context of real-world ion trap setups and significantly reduce anomalous heating of trapped ions scale approximately as $1/h^4$ where $h$ is the ion–electrode distance [14]. This geometrical optimization precedes the adjustment of the Mathieu stability parameters a and q, which are subsequently selected to achieve stable and well-controlled ion confinement.

In this study, the surface-electrode ion trap was optimized to maximize the Trap depth at a predefined ion height. The optimization focused on key geometric parameters, including the ion–electrode distance and electrode widths, to ensure stable confinement of a large number of ions. To improve laser access and reduce motional heating, the ion height was chosen to be 134µm once all the potentials are applied.

In order to achieve required ion height, the radiofrequency (RF) electrodes were designed with widths of approximately 300µm. These electrodes were separated by a central ground/control electrode of width equal to 85µm, as shown in Fig. 1. Along the axial direction, segmented DC electrodes with widths (in axial direction) of about 310µm were incorporated. These DC electrodes provided axial confinement and enabled precise control of the trapping potential. The choice of the RF voltage VRF was guided by several practical and physical constraints. The maximum applicable voltage is limited by the breakdown voltage of the electrodes, which depends on the substrate and insulator materials, electrode spacing, fabrication quality, and vacuum conditions. In addition, RF power dissipation can cause electrode heating and contribute to anomalous heating of the trapped ions. Outgassing from trap materials under high RF fields also places further constraints on the operating voltage. For microfabricated surface traps with electrode gaps larger than 5µm, RF voltages up to 500V can be applied safely [19].

Considering these limitations and the required trap parameters, the initial trapping configuration was chosen as follows. The central electrode was held at ground potential (0V), while the outer RF electrodes were driven with an amplitude of 200V at a frequency Ω= 2 π 22MHz. The DC control electrodes were biased at 6V and -8.4V. Under these conditions, ions can be stably trapped at the targeted height determined by the trap geometry. To shuttle the ion toward the surface, the RF voltage applied to the central electrode(s) gradually increased up to half of $V_{RF}$, while keeping the remaining voltages fixed. As the RF voltage on the central electrode(s) was raised, the local minima shifted downward, causing a controlled displacement of the ion position. In this way, the ion height was reduced from approximately 134µm to 86µm. The progression of the local minima during this process is illustrated in Fig.2, which shows a cross-sectional contour plot of the trapping potential at two cross-sectional trapping regions.

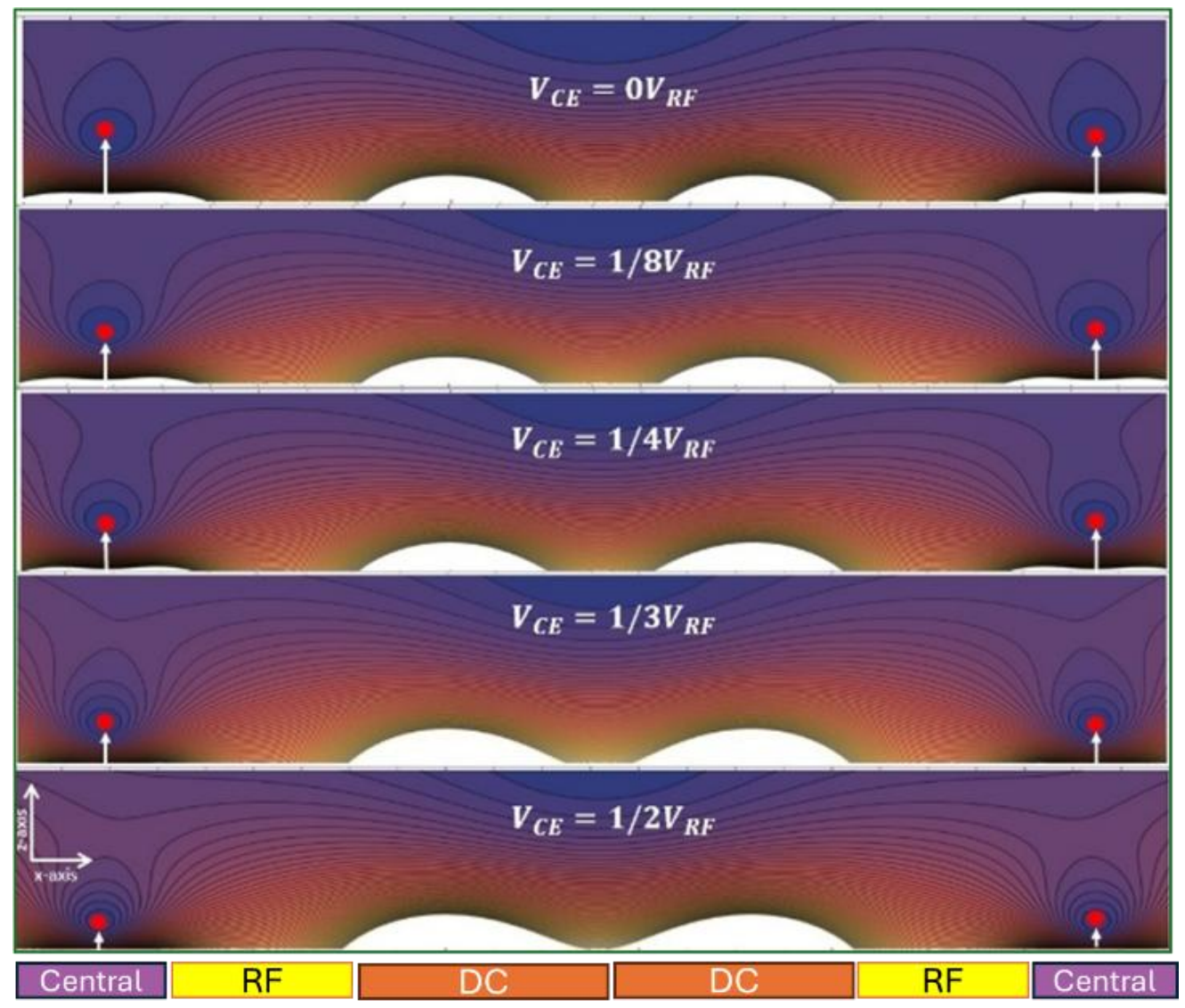

Figure 2. Illustration of the variation in position of the trapping regions over the central electrodes in two trapping zones as the RF voltage $V_{CE}$ is applied gradually to the central electrode.

## III. VERTICAL SHUTTLING

In this work, we adopt a conceptually simple and experimentally robust method demonstrated in [14] . In this scheme, RF voltage is applied to central electrodes shown in Fig. 1(a), which is nominally held at ground potential during standard trapping operation. Following ion loading and confinement, the application of an RF voltage to this central electrode generates an additional pondermotive force that displaces the ion toward the electrode surface. By gradually increasing the RF amplitude on the central electrode, the ion can be shuttled vertically in a controlled downward direction.
To ensure adiabatic transport and suppress excess motional excitation during vertical shuttling, it is necessary to simultaneously adjust the DC voltages applied to the DC control electrodes. This compensation maintains a near-zero electric field along the ion trajectory for each intermediate ion position. Inadequate DC compensation during the shuttling process can introduce stray electric fields, leading to increased motional heating and degradation of trapping performance.

### *A. Variation in trap parameters during verical shuttling*

Owing to the symmetric multi-zone trap design, the vertical shuttling protocol, implemented through globally applied time-dependent RF and DC voltages, operates identically across all trapping zones, thereby inherently incorporating multi-zone functionality and ensuring ion transport dynamics throughout the architecture.

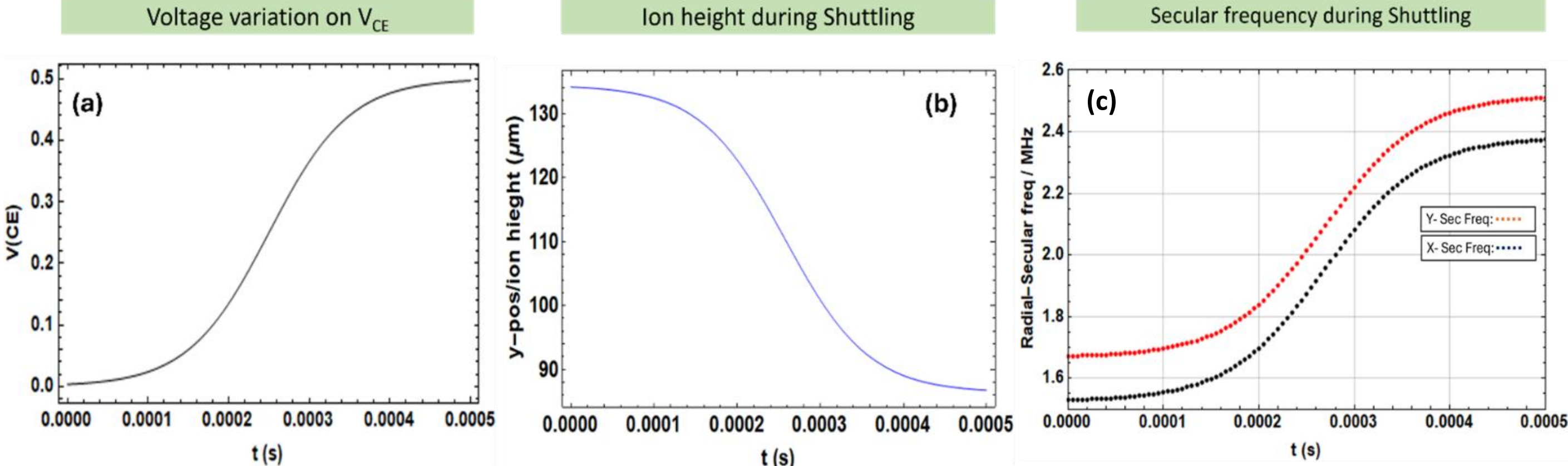

Figure 3. Variation of trap parameters during ion shuttling over the time interval $'t'$, showing when; (a) the RF voltage on the central electrode $V_{\mathrm{CE}}$ is applied. (b) The corresponding displacement of the ion position in vertical direction and (c) the variation of the radial secular frequencies (x and y).

By focusing on a single zone, the influence of vertical shuttling on various trap factors is summarized in Fig.3 and Fig.4 as the voltage on central (ground) electrode ($V_{CE}$) varies causing ion transportation, as shown in Fig.3(a).

### 1. *Ion Height*

During vertical shuttling, the height of the trapped ion was controlled by adjusting the voltage applied to the central ground electrode. When the central electrode was held at 0V, the ion was located at its maximum height of approximately 132µm above the trap surface. As the voltage on this electrode was increased, as shown in Fig. 3(a) the ion height decreased to 86µm in a systematic manner, resulting in a controlled downward displacement of the ion. Consequently, the ion is shuttled vertically downward as the voltage is increased. The reduction in ion height with increasing voltage within time 't' is clearly shown in Fig. 3(b).

### 2. *Secular Frequency*

It is important to note that the radial secular frequencies along both the x- and y-directions evolve in a similar manner during vertical shuttling, as both modes are governed by the same time-dependent RF pseudopotential. Fig. 3(c) illustrates a clear increase in the radial secular frequencies as the ion is transported along the vertical (y) axis over the time interval *t*. Despite this variation, the radial confinement remains within a stable operating regime (1.5—2.5 MHz) and does not adversely affect trapping performance or experimental stability. The axial secular frequency (z-direction) exhibits similar, but comparatively smaller variation, increasing from approximately 300 kHz to 400 kHz as the ion is shuttled downward, closer to the DC electrode. In addition, both radial and axial secular frequencies were independently validated through analysis of the ion's kinetic energy evolution during transport, as presented in Fig. 7 and discussed in the subsequent section.

### 3. *Trap Depth*

As the ion is moved closer to the trap surface, the effective RF pseudopotential becomes stronger, leading to deeper confinement. This behavior can be seen in Fig. 4, which shows the trap depth as a function of ion position during the downward shuttling process induced by increasing the voltage $V_{CE}$ on the central electrode.

## I. VERTCAL SHUTTLING PROTOCOL

The primary goal of an optimal shuttling protocol is to establish a controlled and systematic method for transporting ions to arbitrary positions above the trap surface while preserving the quantum information encoded in their internal and motional states. Although high-fidelity ion shuttling across various ion traps has been demonstrated experimentally, the realization of fully adiabatic shuttling remains challenging [7,8,10]. This limitation arises mainly from motional heating of the ion during the transport process. Nonetheless, ion heating associated with shuttling can be significantly reduced by carefully engineered electrode voltage waveforms. In this context, Hucul *et al.* [17] have developed a comprehensive theoretical framework that provides a detailed description of ion-shuttling dynamics and the mechanisms governing motional excitation during transport.

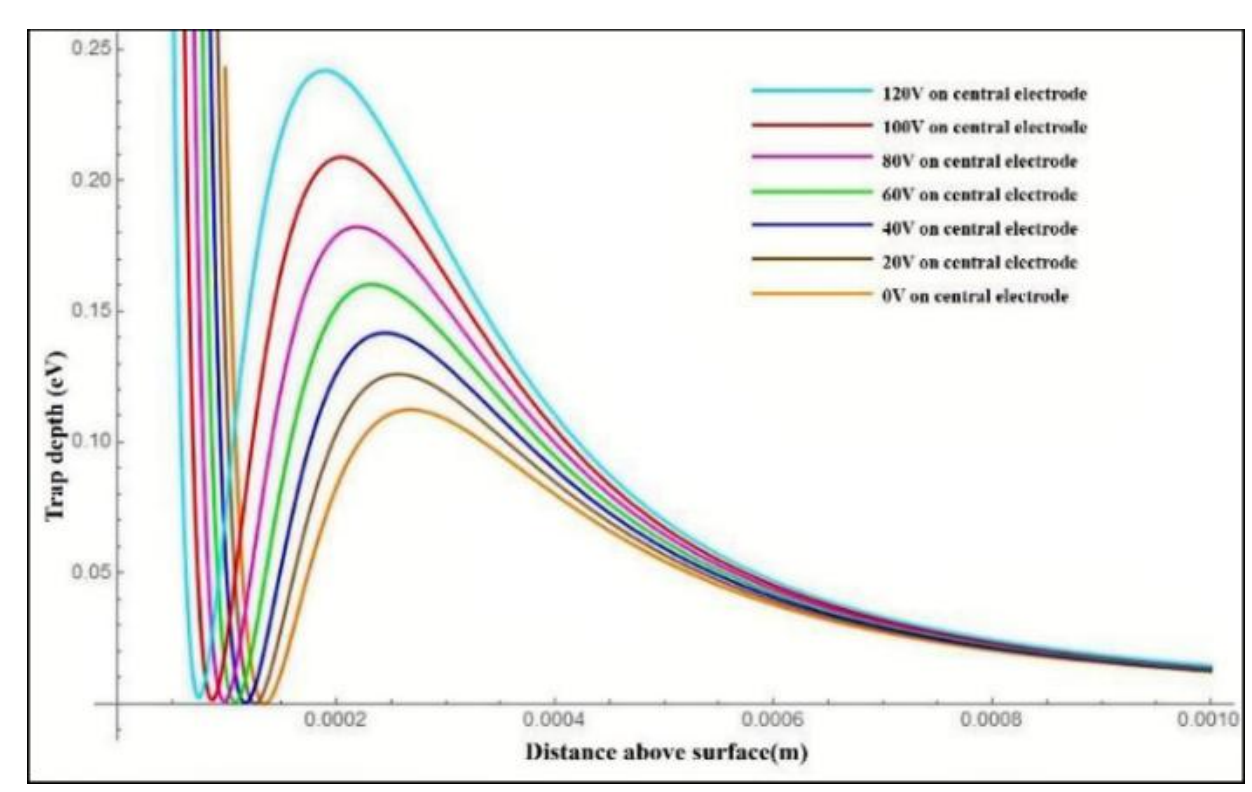

Figure 4. The plots show a significant increase in trap depth from 0.12 eV to 2.4 eV as the ion height is reduced from 132µm to 86µm, achieved by applying $V_{CE}$ to the central electrode.

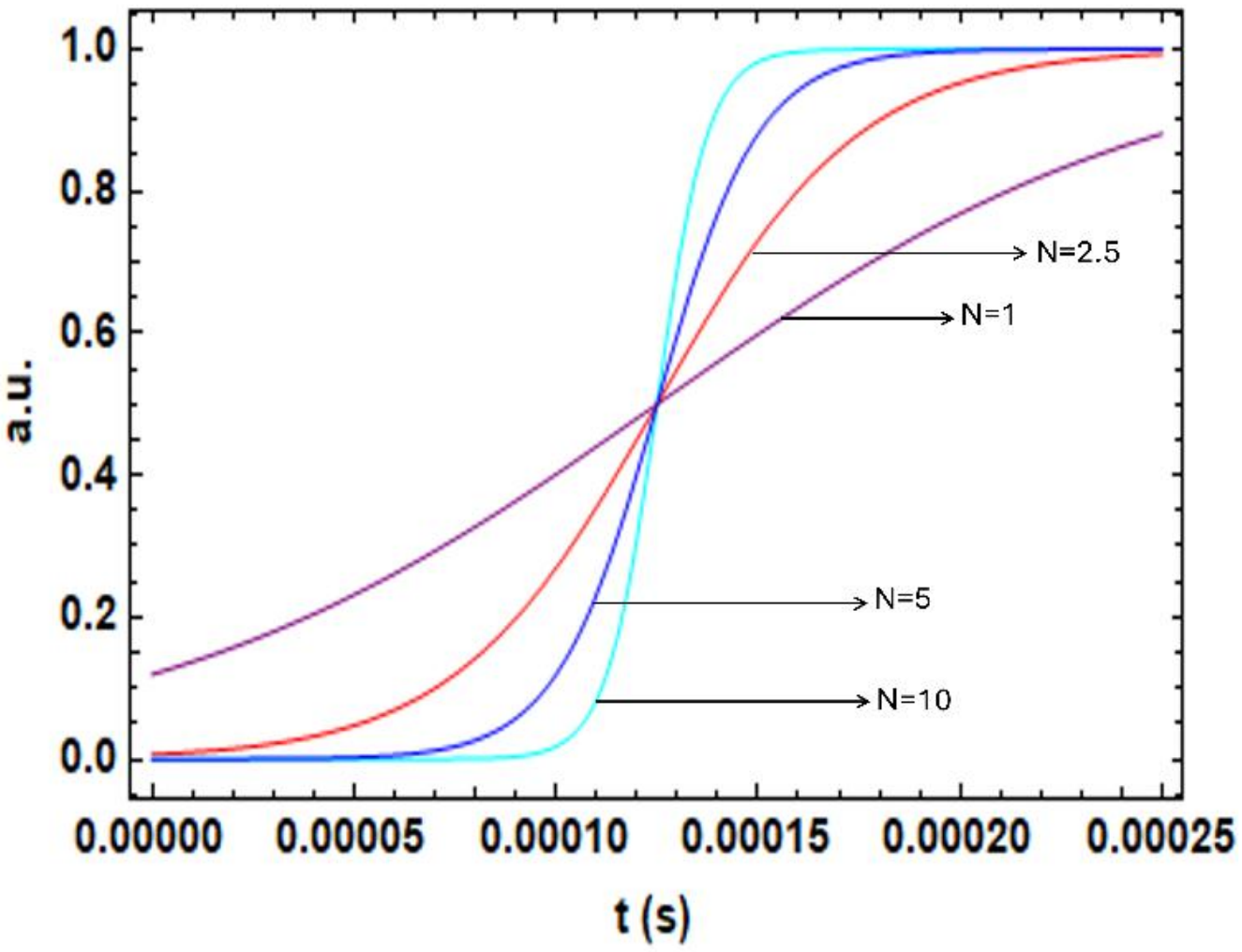

Figure 5. The voltage variation applied to the central electrode follows a hyperbolic tangent function, strongly governed by the parameter *N*; relatively smaller values of *N* lead to a near-linear transport trajectory. The function is presented in arbitrary units.

### *A. Protocol for smooth shuttling of ion*

Hucul *et al.* [17] emphasized the importance of inertial control of ions during the initiation and termination phases of linear shuttling operations. They introduced several strategies aimed at suppressing excessive motional energy gain during linear transport. A practical and effective approach involves gradually varying the voltage applied to the central RF electrode over a defined time interval. This controlled voltage ramp induces a smooth displacement of the trapping potential minimum along the ion trajectory, enabling transport from the initial position to the desired final location. In their study, they performed a comparative analysis of three distinct shuttling trajectories: a linear trajectory $x_{0L}(t)$, a sinusoidal trajectory $x_{0S}(t)$, and a hyperbolic tangent trajectory $x_{0T}(t)$. Owing to the vertical nature of the shuttling process, in this study, we consider only the $y$-coordinate. Among these options, the hyperbolic tangent trajectory was identified as the most favorable, as illustrated in Fig. 5. The trajectory is determined using:

$$y_{oT}(t) = \frac{L}{2}\frac{\tanh\left(\frac{N(2t-T)}{T}+\tanh(N)\right)}{\tanh(N)}\big(H(t)-H(t-T)+LH(t-T)\big), \qquad (2)$$

In these expressions, $L$ denotes the total distance covered during the shuttling operation, $T$ represents the total duration of the shuttle, and $H$ is the Heaviside step function. In the case of the hyperbolic tangent trajectory, the parameter $N$ plays a crucial role by controlling the smoothness of the velocity profile of the potential minimum at the beginning and end of the shuttling protocol.

### *B. RF and DC electrode volatge variation for ion shuttling*

To achieve precise control of ion motion during linear or vertical transport, it is essential to generate accurate, time-dependent electric fields. These fields serve as the fundamental control primitives and are implemented experimentally by applying appropriately timed voltage waveforms to the trap electrodes. The required electrode voltages are obtained by scaling these base control operations in accordance with the desired ion trajectory.

During the ion shuttling, the RF voltage applied to the central electrode is ramped up from 0V to 100V, while the negative DC electrode voltage is changed slightly from −8.4V to −8.25V according to Eq.3.

$$V = b_1[t]\frac{(a_2-a_1)}{2}+\frac{(a_1+a_2)}{2} \qquad (3)$$

Here, $a_1$ and $a_2$ denote the initial and final values of the applied voltages, respectively. The function $b_1[t] = Tanh(\frac{t-t_1}{T})$ represents a time-dependent hyperbolic tangent step function, where $t_1$ is the shuttling time and $T$ determines the steepness of the voltage transition. Whereas RF voltage on RF electrodes, is kept constant at 200V and DC voltage on endcap electrodes kept at +6V. Fig. 1(a) shows the corresponding position of the electrodes. This combined RF and DC voltage control enables stable and low-heating vertical ion transport.

### *C. Shuttling Profile*

The transport protocol begins with calculating the electric potential generated by the trap electrodes during shuttling process. The resulting ion dynamics are then evaluated by solving the differential equations of motion numerically using *NDSolve* function in *Mathematica*, from which the motional energy gained by the ion during transport can be directly determined.

A typical vertical shuttling profile with *N*=10 is shown in Fig.6, with voltage variation on central electrode $V_{CE}$ (Fig.6a) where the reduction of height $h$ (Fig.6b) can be seen as the RF voltage $V_{CE}$ applied to the central RF electrode is increased from 0V to 0.5 of $V_{RF}$ over a time interval of 1ms following the *N*=10. The profile indicates a rapid initial acceleration of the ion during the early stage of the shuttling process, which leads to a maximum in the kinetic energy gain, can be seen in (Fig.6c). Whereas in Fig. 6(d), the voltage variation profile is overlaid with the ion shuttling trajectory over the given time interval, illustrating their direct correlation during the transport process.

As a consequence of this acceleration, the trapped ion acquires additional motional energy, quantified in terms of the mean motional quanta $\langle n \rangle$, which can be calculated using following Eq.4 and 5;

$$\langle n \rangle = \frac{\frac{1}{2}m(\nu - \nu_0(t)^2) + \frac{1}{2}m\omega_t^2(x - x_0(t))^2}{\hbar\omega} \tag{4}$$

and during the shuttling process, as

$$\langle n \rangle_s = \frac{\text{Final } K.E_{max} - \text{Initial } K.E_{max}}{\hbar\omega} \tag{5}$$

The shuttling trajectory therefore highlights the direct correlation between the applied RF voltage ramp on the central electrode, the resulting ion displacement, and the motional excitation induced during transport is maximum at the mid of the shuttling protocol as seen in Fig 7.

## II. Gain of Kinetic energy and Motional Quanta

### A. *Impact of N parameter*

Abrupt accelerations experienced by the trapped ion at the beginning and end of the shuttling process can lead to excessive motional energy gain, thereby increasing the risk of decoherence. To suppress this unwanted kinetic energy growth during transport, a hyperbolic tangent (*Tanh*) voltage profile is employed with an appropriately chosen value of the parameter $N$. This parameter controls the rate at which the hyperbolic tangent function varies in time and therefore determines the smoothness of the shuttling trajectory.
As the duration of the shuttling process is increased, a clear dependence of the kinetic energy gain on the parameter $N$ becomes evident. As shown in Fig. 8, longer shuttling times lead to a systematic reduction in the motional energy gained by the ion. However, for a fixed shuttling duration, the kinetic energy gain increases significantly with increasing $N$. In particular, the kinetic energy gain is observed to increase by approximately a factor of four when the parameter $N$ of the hyperbolic tangent profile is doubled, as illustrated in the kinetic energy versus time plot.

More generally, if the parameter $N$is scaled by a factor $m$, with $m = 1,2,3, \ldots$, the resulting kinetic energy gain approximately follows a quadratic dependence, $\mathrm{KE}_{\mathrm{gain}} \propto m^2$. Summarized data can be used to concisely present the dependence of kinetic energy gain on the choice of $N$and the shuttling duration, providing a clear overview of the trade-offs involved in optimizing the shuttling protocol.

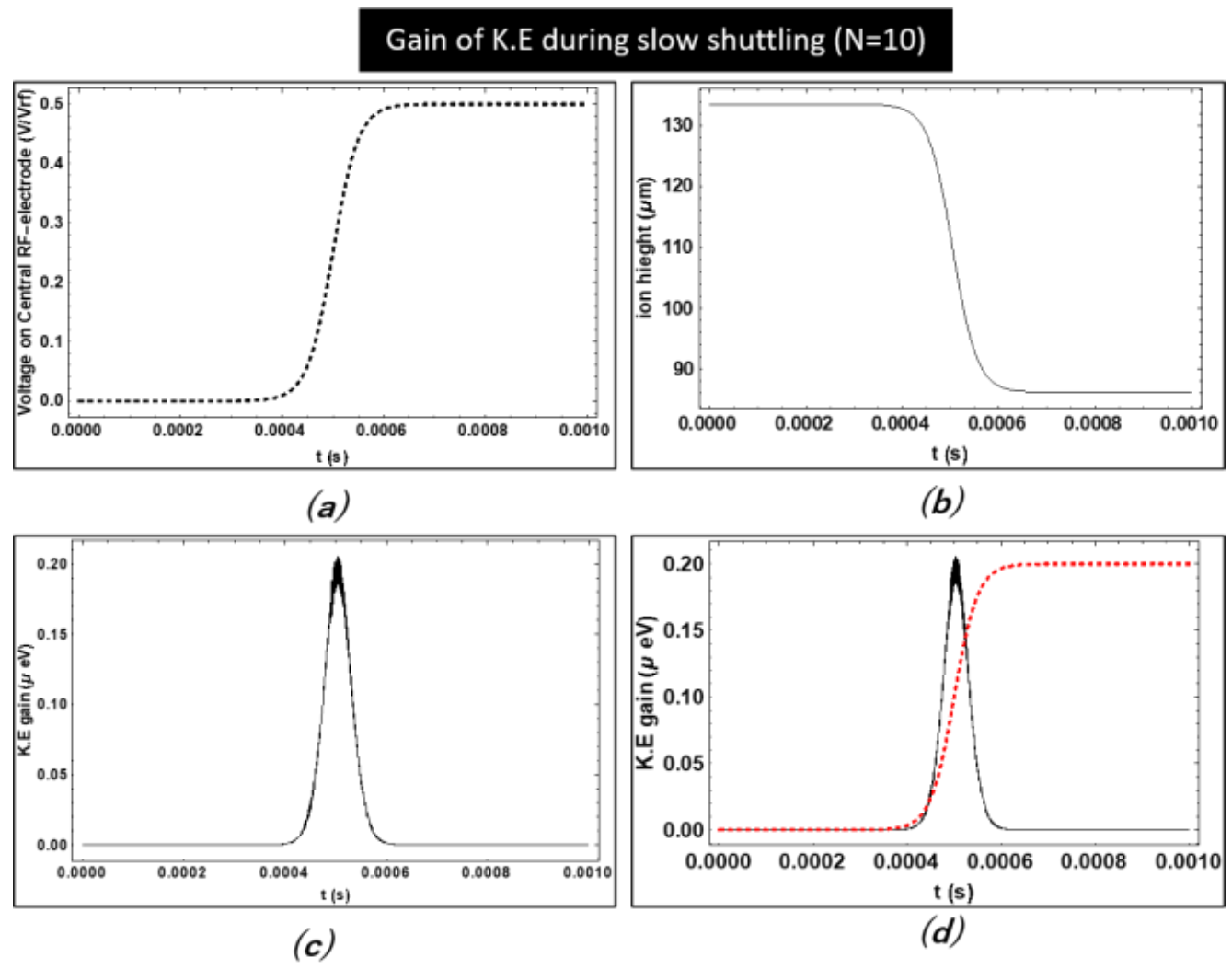

Figure 6. Ion a shuttling profile with 1ms time scale and *N*=10. (a) when voltage was increased from 0V to 100V on central RF electrode during 1ms, (b) the ion is decreased from 134μm to 86μm, (c) the gain of K.E as the ion will experience a kick during shuttling so the gain is maximum (d) overlap of (a and c) plots show sharp rise of energy gain at voltage change duration.

### D. *Anamlous heating rate*

Experiments have shown that measured heating rates in trapped-ion systems are significantly higher than those expected from thermal electron motion in conductors. This excess heating, commonly referred to as *anomalous heating*, is not yet fully understood, as its microscopic origin remains unclear. Comparative studies of ion traps with different geometries and length scales have consistently demonstrated that the anomalous heating rate increases strongly as the distance between the ion and the electrode surface is reduced.

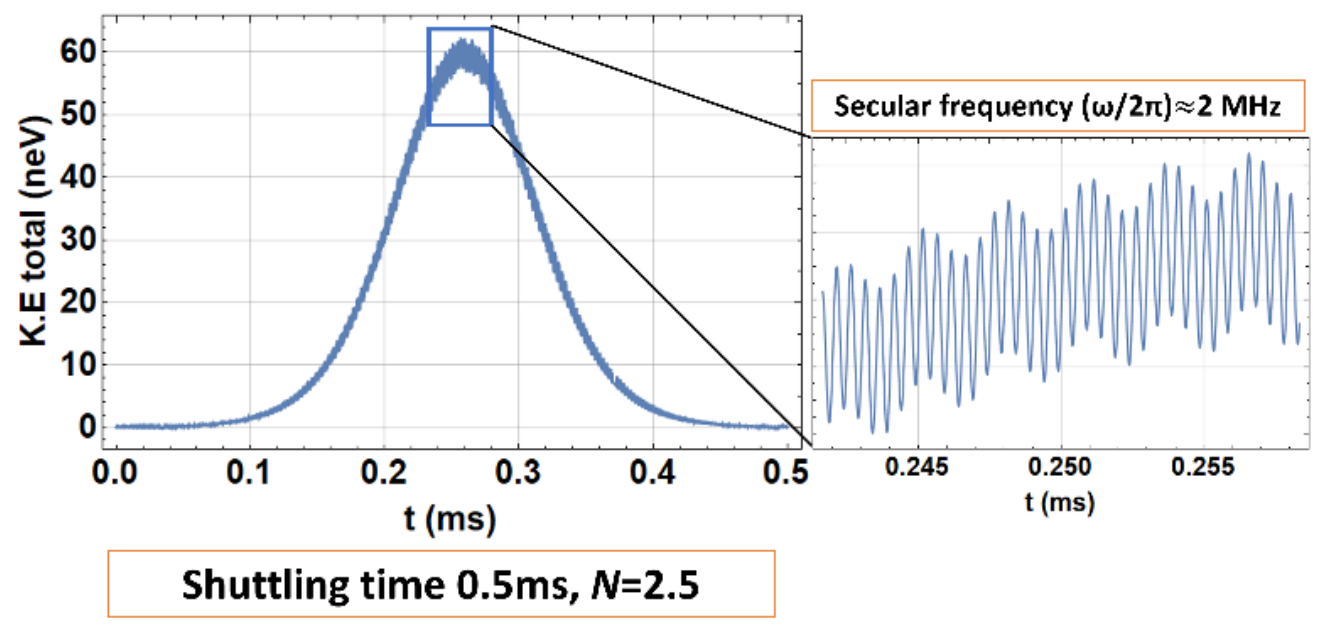

Figure 7. The maximum motional excitation induced during vertical transportation of an ion with *N* parameter of 2.5 and shuttling time of 0.5ms. The zoomed-in section illustrates the evolution of the secular frequencies during transport: the radial secular frequency remains approximately ~2MHz, while the beat oscillations correspond to an axial secular frequency of ~375kHz.

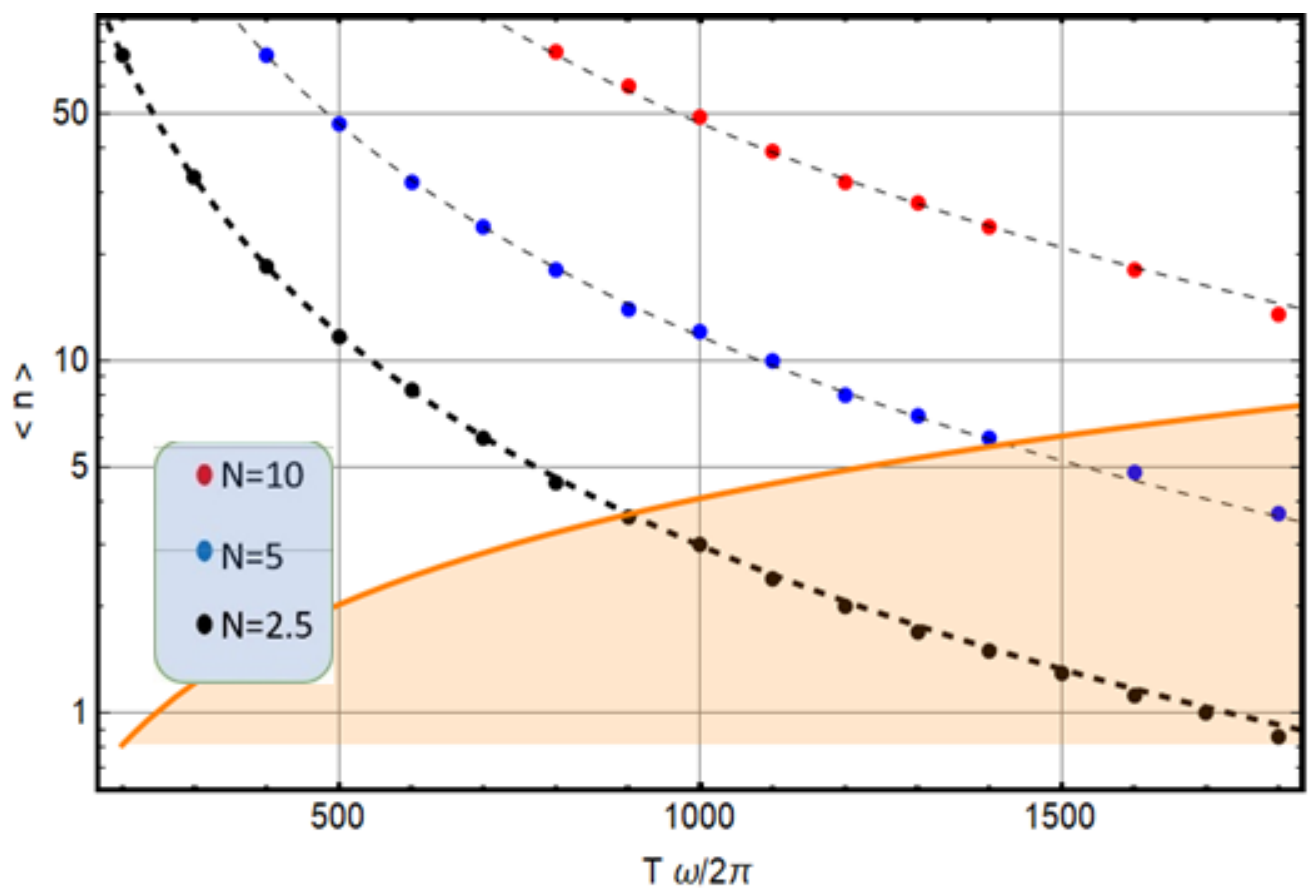

Figure 8. Comparison of motional excitation <n> versus shuttling time *t* (in cycles) for different *N* parameters. The orange curve shows anomalous heating, with the shaded area indicating accumulated motional gain. Notably, *N*=2.5 yields the minimum net motional energy gain among the considered cases

Consequently, during vertical shuttling, where the ion–electrode separation decreases, the anomalous heating rate rises sharply.

For planar surface-electrode traps, the heating rate at an ion–electrode distance of 134 ± 1.5 μm has been measured to be 3.1 ± 0.35 quanta/ms and is found to scale approximately as $1/h^4$ where h is the ion–electrode distance [14]. The combined energy gain arising from anomalous heating and shuttling-induced excitation, plotted as a function of the number of shuttling cycles with the horizontal axis normalized to the secular frequency (number of cycles equal to $T\,\omega/2\pi$), is shown in Fig.8.

Assuming a linear shuttling trajectory with a hyperbolic tangent parameter *N* equal to 2.5, the motional heating of the ion due to anomalous effects during transport can be calculated over the shuttling time T using Eq.6.

$$Q = \int_0^T \frac{k}{[r(t)]^4} dt \qquad (6)$$

The average heating rate during vertical shuttling from an ion–electrode distance, for example, of 134 micrometers to 86 micrometers over a shuttling duration of 0.5 milliseconds, with the scaling factor $k = 3 \times (134)^4$, is estimated to be approximately 8.14 motional quanta. The optimal operating point corresponds to the condition where the anomalous heating contribution intersects with the motional excitation arising from the shuttling process itself.

For a hyperbolic tangent parameter value of $N = 2.5$, this intersection occurs at a mean motional occupation number of approximately ⟨n⟩ equal to 3.5 after about 900 shuttling cycles. Dividing this cycle count by the axial secular frequency of approximately 2 MHz yields a corresponding shuttling time of about 0.45ms.

## III. DISCUSSION ON RESULTS/ CONCLUSION

The observed behavior of motional excitation during vertical ion shuttling confirms that, for a fixed shuttling distance, the gain in motional quanta of the transported ion is strongly dependent on the parameter $N$ of the hyperbolic tangent voltage profile. Larger values of $N$ lead to a more rapid variation of the shuttling potential, resulting in a higher motional excitation over shorter time intervals. In contrast, smaller $N$ values produce smoother trajectories and therefore yield a reduced number of motional quanta for short shuttling durations. Consequently, selecting an optimal value of $N$ that minimizes the motional excitation $\langle n \rangle$ while maintaining a short transport time is essential for efficient ion shuttling.

Based on this optimization criterion, a value of $N = 2.5$ is identified as optimal for the vertical shuttling protocol driven by the central RF electrode. At a shuttling time of approximately 0.45 milliseconds, the total motional excitation is limited to about ≈7 quanta. These results demonstrate that careful optimization of the hyperbolic tangent profile parameter enables fast and efficient vertical ion shuttling with minimal motional excitation, thereby supporting high-fidelity transport compatible with scalable quantum information processing and sensing applications.